\RequirePackage{lineno}
\documentclass[twocolumn,showpacs,amsmath,amssymb,superscriptaddress,nofootinbib]{revtex4-1}
\usepackage{amsmath}
\usepackage{amssymb}
\usepackage{graphicx}
\usepackage{dcolumn}
\usepackage{bm}

\usepackage{amsfonts}
\usepackage{keyval,graphicx}
\usepackage{textcomp,wasysym}
\usepackage{multirow}

\usepackage[dvipdfm,CJKbookmarks=true,unicode,colorlinks,linkcolor=blue,anchorcolor=blue,citecolor=blue,pdfborder={0 0 0}]{hyperref}

\newcommand{\jpsi}{J/\psi}
\newcommand{\g}{\gamma}
\newcommand{\pip}{\pi^+}
\newcommand{\pim}{\pi^-}
\newcommand{\piz}{\pi^{0}}
\newcommand{\ks}{K^0_S}
\newcommand{\fz}{f_{0}(980)}
\newcommand{\etap}{\eta^{\prime}}
\newcommand{\pbar}{\bar{p}}
\newcommand{\kap}{K^+}
\newcommand{\kam}{K^-}
\newcommand{\ar}{\rightarrow}
\newcommand{\GeV}{GeV/$c^2$}
\newcommand{\MeV}{MeV/$c^2$}

\begin{document}

\title{\boldmath Observation and Spin-Parity Determination of the $X(1835)$ in $\jpsi\ar\g\ks\ks\eta$}

\author{
\small
      M.~Ablikim$^{1}$, M.~N.~Achasov$^{9,f}$, X.~C.~Ai$^{1}$,
      O.~Albayrak$^{5}$, M.~Albrecht$^{4}$, D.~J.~Ambrose$^{44}$,
      A.~Amoroso$^{48A,48C}$, F.~F.~An$^{1}$, Q.~An$^{45,a}$,
      J.~Z.~Bai$^{1}$
% ...
      \\
      \vspace{0.2cm}
      (BESIII Collaboration)\\
      \vspace{0.2cm} {\it
        $^{1}$ Institute of High Energy Physics, Beijing 100049, People's Republic of China\\
% ...
$^{4}$ Bochum Ruhr-University, D-44780 Bochum, Germany\\
        $^{5}$ Carnegie Mellon University, Pittsburgh, Pennsylvania 15213, USA\\
% ...
$^{9}$ G.I. Budker Institute of Nuclear Physics SB RAS (BINP), Novosibirsk 630090, Russia\\
% ...
$^{44}$ University of Rochester, Rochester, New York 14627, USA\\
        $^{45}$ University of Science and Technology of China, Hefei 230026, People's Republic of China\\
% ...
$^{48}$ (A)University of Turin, I-10125, Turin, Italy; (B)University of Eastern Piedmont, I-15121, Alessandria, Italy; (C)INFN, I-10125, Turin, Italy\\
% ...
$^{a}$ Also at State Key Laboratory of Particle Detection and Electronics, Beijing 100049, Hefei 230026, People's Republic of China\\
% ...
$^{f}$ Also at the Novosibirsk State University, Novosibirsk, 630090, Russia\\
% ...
      }
}
\begin{abstract}
We report an observation of the process $\jpsi\ar\g X(1835)\ar\g\ks\ks\eta$
at low $\ks\ks$ mass with a statistical significance
larger than 12.9$\sigma$ using a data sample of $1.31 \times 10^{9}$
$\jpsi$ events collected with the BESIII detector.
In this region of phase space the $\ks\ks$ system is dominantly
produced through the $f_0(980)$.
By performing a partial wave analysis, we determine the spin parity of
the $X(1835)$ to be $J^{PC}=0^{-+}$. The mass and width of the
observed $X(1835)$ are $1844\pm9(\text{stat})^{+16}_{-25}(\text{syst})$~\MeV~and
$192^{+20}_{-17}(\text{stat})^{+62}_{-43}(\text{syst})$~MeV, respectively, which are
consistent with the results obtained by BESIII in the channel $\jpsi\ar\g\pip\pim\etap$.
\end{abstract}


\maketitle

The non-Abelian property of quantum chromodynamics (QCD) permits the
existence of bound states beyond conventional mesons and baryons, such
as glueballs, hybrid states, and multiquark states. The search for
these unconventional states is one of the main interests in
experimental particle physics. One of the most promising
candidates, the $X(1835)$ resonance, was first observed in its decay to
$\pip\pim\etap$ in the process $\jpsi\ar\g\pip\pim\etap$ by
BESII~\cite{x1835_bes2}; this observation was subsequently confirmed by
BESIII~\cite{x1835_bes3}. The discovery of the
$X(1835)$ has stimulated theoretical speculation concerning its
nature. Possible interpretations include a $p\pbar$ bound
state~\cite{x1835theory_baryonium}, a second radial excitation of the
$\etap$~\cite{x1835theory_etapexcitation}, and a pseudoscalar
glueball~\cite{x1835theory_glueball}. In addition, an enhancement in
the invariant $p\pbar$ mass at threshold, $X(p\pbar)$, was first
observed by BESII in the decay $\jpsi\ar\g p\pbar$~\cite{gppb_bes2},
and was later also seen by BESIII~\cite{gppb_bes3} and
CLEO~\cite{gppb_cleo}.
In a partial-wave analysis of $\jpsi\ar\g p\pbar$, BESIII determined
the $J^{PC}$ of the $X(p\pbar)$ to be $0^{-+}$~\cite{gppbpwa_bes3}.
The
mass of the $X(p\pbar)$ is consistent with the $X(1835)$ mass measured
in $\jpsi\ar\g\pip\pim\etap$~\cite{x1835_bes3}, but the width of the
$X(p\pbar)$ is significantly narrower.

To understand the nature of the $X(1835)$, it is crucial to measure
its $J^{PC}$ and to search for new decay modes. Because of its
similarity to $\jpsi\ar\g\pip\pim\etap$, $\jpsi\ar\g K\bar{K}\eta$ is
a favorable channel to search for $X(1835)\ar K\bar{K}\eta$.  In contrast
to $\jpsi\ar\g\kap\kam\eta$, there is no background contamination for
$\jpsi\ar\g\ks\ks\eta$ from $\jpsi\ar\piz\ks\ks\eta$ and
$\jpsi\ar\ks\ks\eta$, which are forbidden by exchange symmetry and
$CP$ conservation. Therefore, the channel $\jpsi\ar\g\ks\ks\eta$
provides a clean environment with minimal uncertainties due to
background modeling. In this Letter, we report the first observation
and spin-parity determination of the $X(1835)$ in
$\jpsi\ar\g\ks\ks\eta$, where the $\ks$ and $\eta$ are reconstructed
from their decays to $\pip\pim$ and $\gamma\gamma$, respectively. The
analysis is based on a sample of $(1310.6\pm10.5) \times 10^{6}$
$\jpsi$ events~\cite{njpsi_09,njpsi_all} collected with the BESIII
detector~\cite{bes3detector}.

The BESIII detector is a magnetic spectrometer operating at BEPCII, a
double-ring $e^+e^-$ collider with center of mass energies between 2.0
and 4.6~GeV. The cylindrical core of the BESIII detector consists of a
helium-based main drift chamber (MDC), a plastic scintillator
time-of-flight system, and a CsI(Tl) electromagnetic calorimeter
(EMC) that are all enclosed in a superconducting solenoidal magnet
providing a 1.0 T (0.9 T in 2012, for about $1087\times 10^6$ collected $J/\psi$)
magnetic field. The solenoid is
supported by an octagonal flux-return yoke with resistive plate
counter muon identifier modules interleaved with steel. The acceptance
of charged particles and photons is 93\% of the 4$\pi$ solid angle,
and the charged-particle momentum resolution at 1~GeV/$c$ is
0.5\%. The EMC measures photon energies with a resolution of
2.5\% (5\%) at 1~GeV in the barrel (end caps). A {\sc
  geant4}-based~\cite{geant4} Monte Carlo (MC) simulation software
package is used to optimize the event selection criteria, estimate
backgrounds, and determine the detection efficiency.

Charged tracks are reconstructed using hits in the MDC. Because there
are two $\ks$ with displaced vertices, the point of closest approach
of each charged track to the $e^{+}e^{-}$ interaction point is
required to be within $\pm30$ cm in the beam direction and within
$40$ cm in the plane perpendicular to the beam direction. The polar
angle between the direction of a charged track and the beam direction
must satisfy $|\cos\theta|<0.93$. Photon candidates are selected
from showers in the EMC with the energy deposited in the EMC barrel
region $(|\cos \theta|<0.8)$ and the EMC end caps region
$(0.86<|\cos \theta|<0.92)$ greater than 25~MeV and 50~MeV,
respectively. EMC cluster timing requirements are used to suppress
electronic noise and energy deposits unrelated to the event.

Candidate $\jpsi\ar\g\ks\ks\eta$ events are required to have four
charged tracks with zero net charge and at least three photon candidates. All
charged tracks are reconstructed under the pion hypothesis. To
reconstruct $\ks$ candidates, the tracks of each $\pip\pim$ pair are
fitted to a common vertex. $\ks$ candidates are required to satisfy
$|M_{\pip\pim}-m_{\ks}|<0.009$~\GeV~and $L/\sigma_{L}>2$, where $L$
and $\sigma_{L}$ are the distance between the common vertex of the $\pip\pim$
pair and the primary vertex, and its error, respectively. The
$\g\g\g\ks\ks$ candidates are subject to a kinematic fit with four
constraints ($4C$), ensuring energy and momentum conservation. Only
candidates where the fit yields a $\chi^{2}_{4C}$ value less than 40
are retained for further analysis. For events with more than three
photon candidates, multiple $\jpsi\ar\g\ks\ks\eta$ candidates are
possible. Only the combination yielding the smallest $\chi^{2}_{4C}$
is retained for further analysis. Candidate $\jpsi\ar\g\ks\ks\eta$
events are required to have exactly one pair of photons within the
$\eta$ mass window ($0.51<M_{\g\g}<0.57$~\GeV). Simulation studies show this
criterion significantly reduces the miscombination of photons from
3.20\% to 0.16\%. The miscombination of pions is also studied and found
to be negligible. To further suppress background events containing a
$\piz$, events with any photon pair within a $\piz$ mass window
($0.10<M_{\g\g}<0.16$~\GeV) are rejected. The decay
$\jpsi\ar\phi\ks\ks$ with $\phi\ar\g\eta$ leads to the same final
state as the investigated reaction $\jpsi\ar\g\ks\ks\eta$. Therefore,
events in the mass region $|M_{\g\eta}-m_{\phi}|<0.04$~\GeV~are
rejected.

\begin{figure}[htpb]
  \centering
    \includegraphics[width=1.6in,height=1.6in]{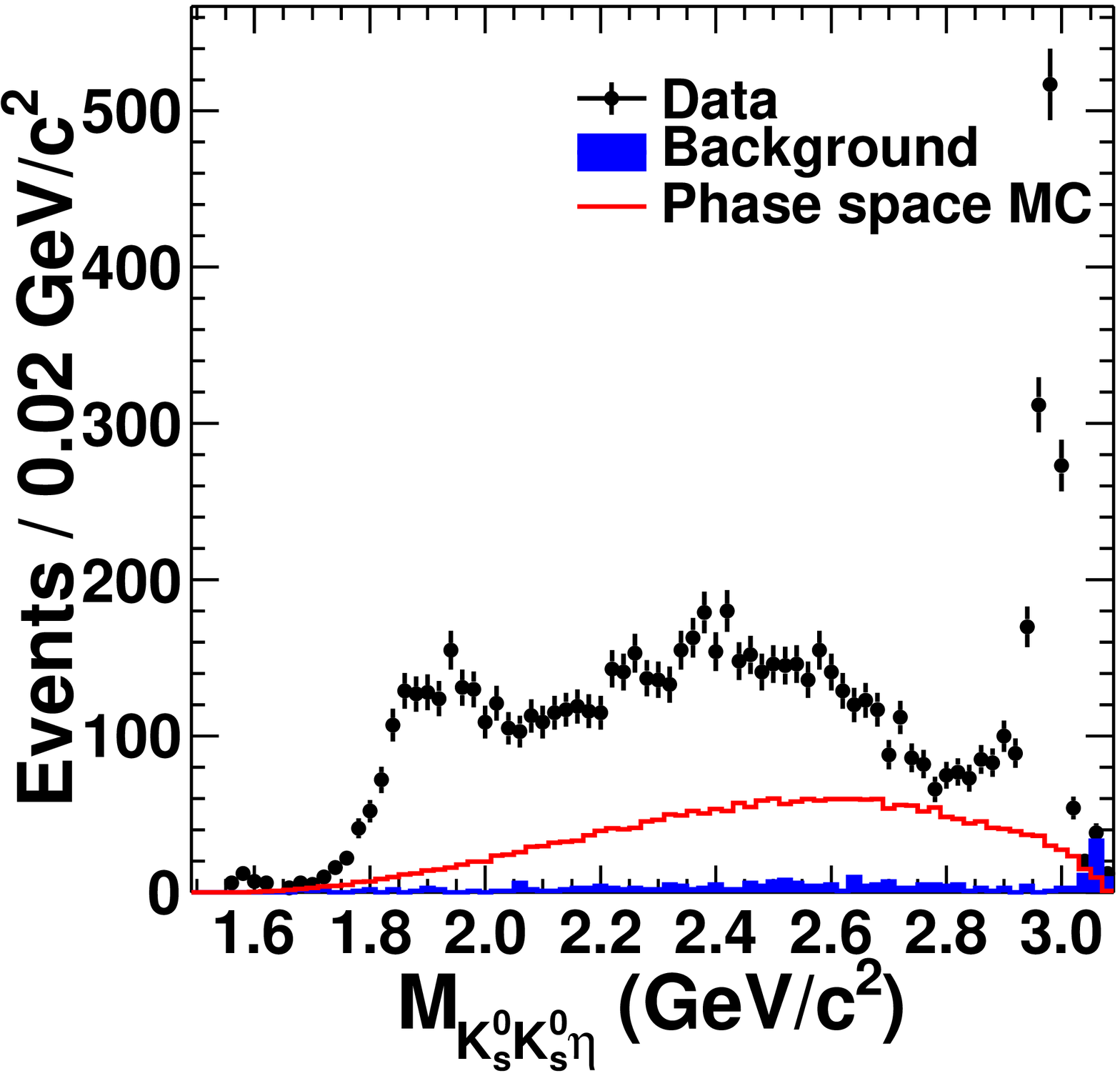}
    \includegraphics[width=1.6in,height=1.6in]{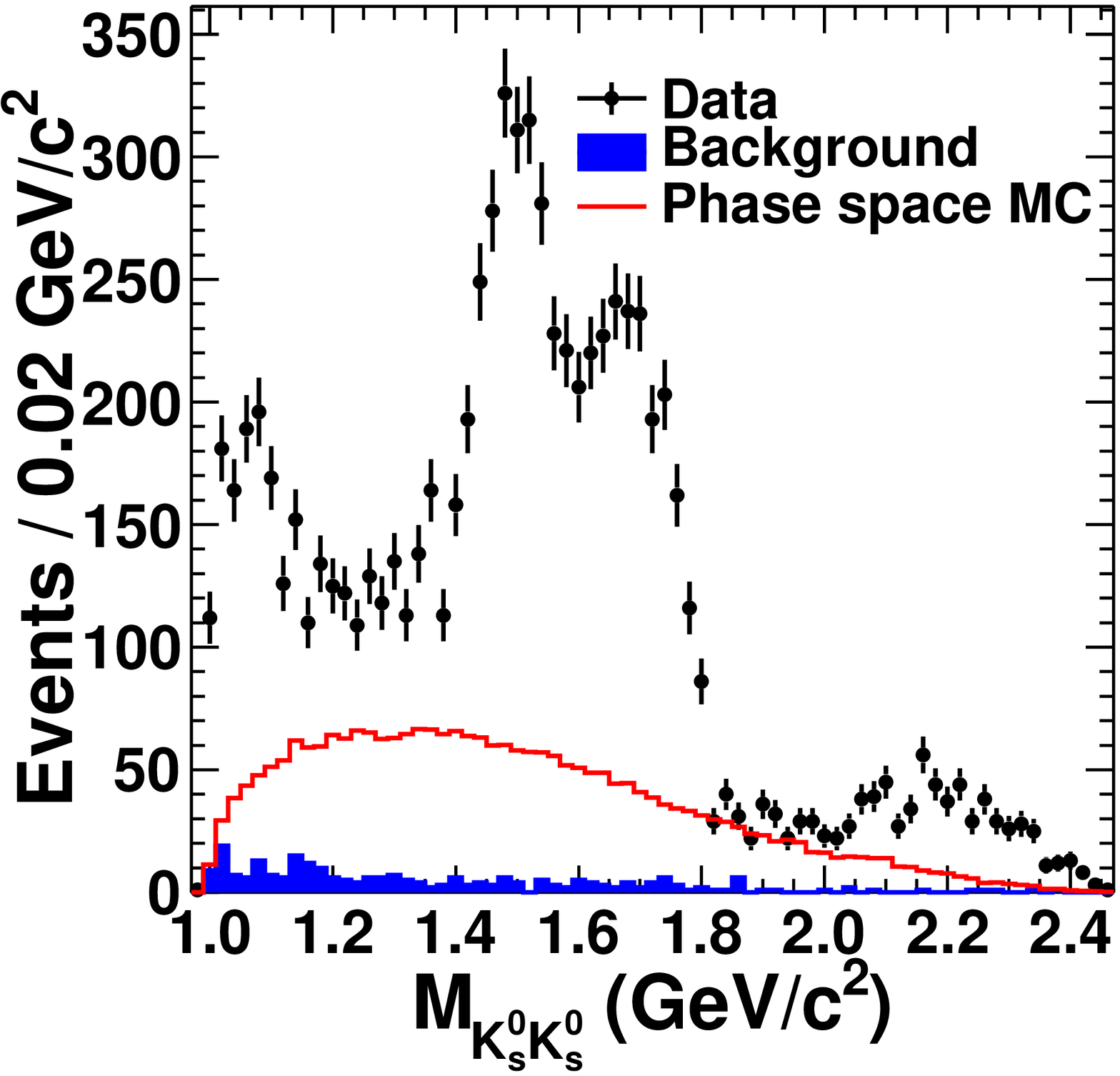}
    \put(-210,105){(a)}
    \put(-90,105){(b)}
    \newline
    \includegraphics[width=1.6in,height=1.6in]{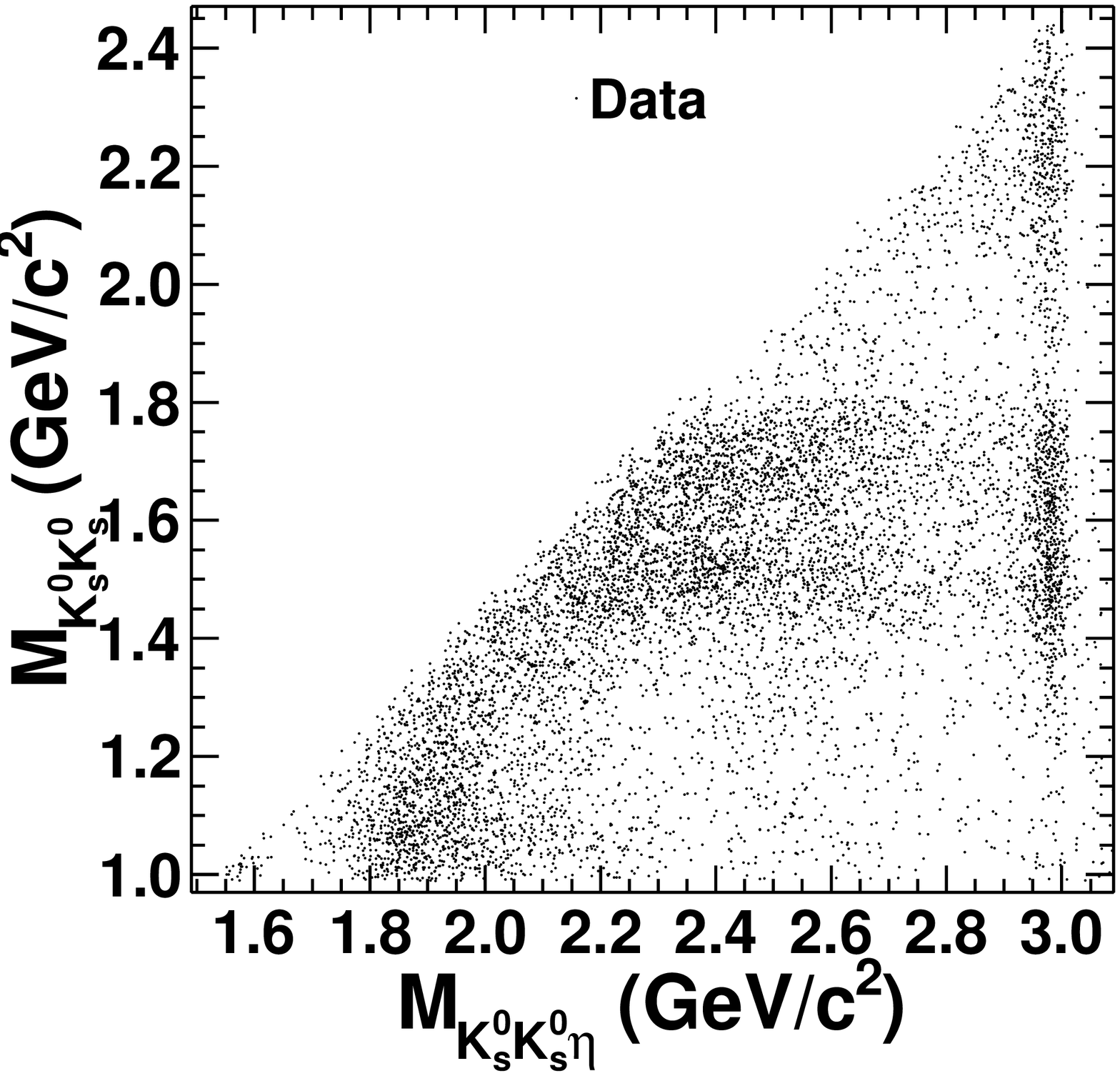}
    \includegraphics[width=1.6in,height=1.6in]{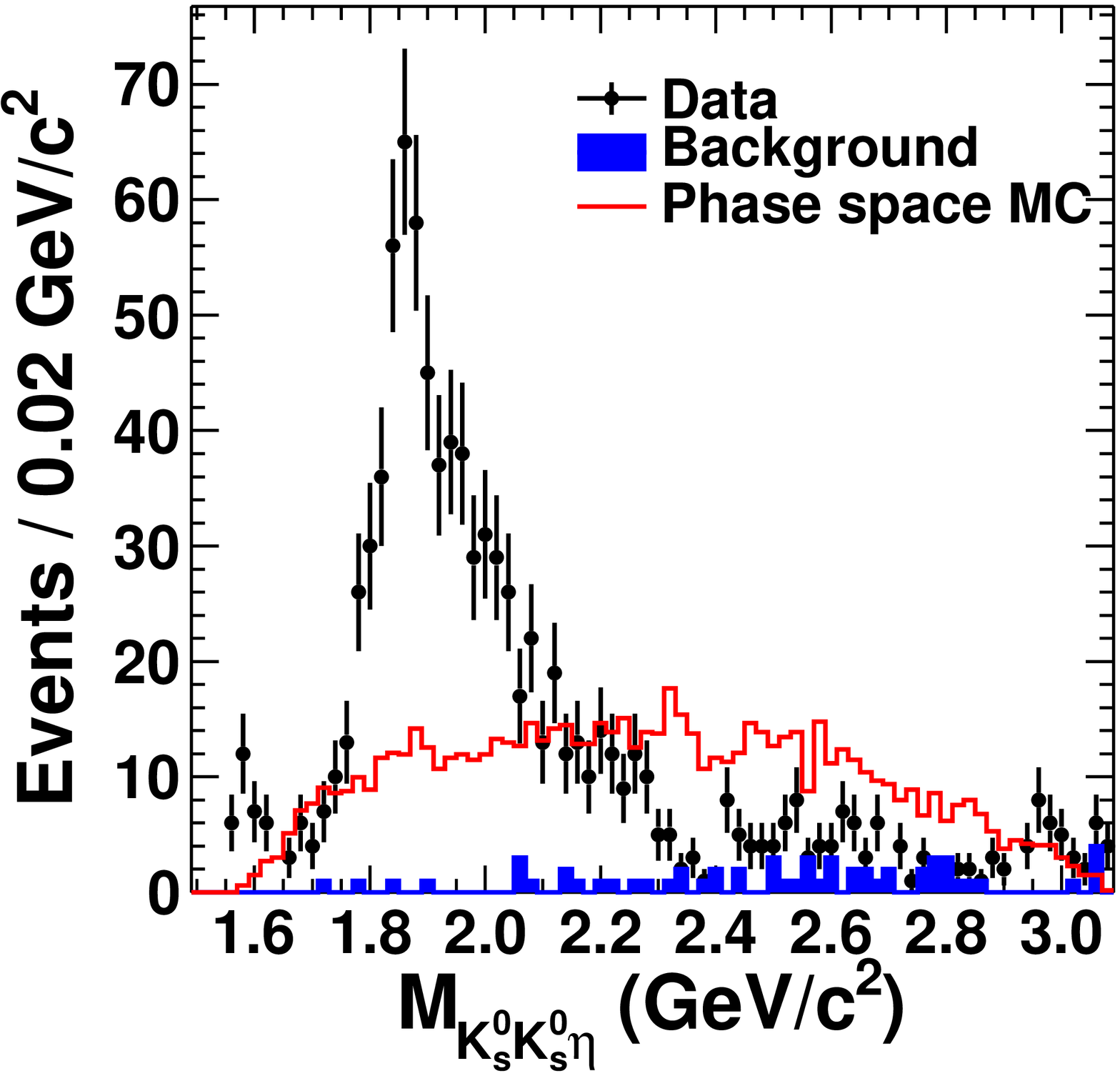}
    \put(-210,105){(c)}
    \put(-90,105){(d)}
    \caption{\label{massdis_dataMCsb}Invariant mass distributions for
      selected events:
      Invariant mass spectra of (a) $\ks\ks\eta$ and (b)  $\ks\ks$;
      (c) scatter plot of $M_{\ks\ks}$ versus $M_{\ks\ks\eta}$;
      (d) $\ks\ks\eta$ invariant mass spectrum for events with
      the requirement $M_{\ks\ks}<1.1$~\GeV. Dots with error bars
      are data; the shaded histograms are the non-$\eta$ backgrounds
      estimated by the $\eta$ sideband; the solid histograms are phase
      space MC events of $\jpsi\ar\g\ks\ks\eta$ with arbitrary
      normalization.}
\end{figure}

After applying the selection criteria discussed above, the invariant
mass spectrum of $\ks\ks\eta$ shown in Fig.~\ref{massdis_dataMCsb} (a)
is obtained. Besides a distinct $\eta_{c}$ signal, a clear structure
around 1.85~\GeV~is observed. The $\ks\ks$ mass spectrum, shown in
Fig.~\ref{massdis_dataMCsb} (b), reveals a strong enhancement near the
$\ks\ks$ mass threshold, which is interpreted as the $f_{0}(980)$ by
considering spin-parity and isospin conservation. The scatter plot of
the invariant mass of $\ks\ks$ versus that of $\ks\ks\eta$ is shown in
Fig.~\ref{massdis_dataMCsb} (c). A clear accumulation of events is
seen around the intersection of the $\fz$ and the structure around
1.85~\GeV. This indicates that the structure around 1.85~\GeV~is
strongly correlated with $\fz$. By requiring $M_{\ks\ks}<1.1$~\GeV, the
structure around 1.85~\GeV~becomes much more prominent in the
$\ks\ks\eta$ mass spectrum [Fig.~\ref{massdis_dataMCsb} (d)]. In
addition, there is an excess of events around 1.6~\GeV.

Potential background processes are studied using a simulated sample of
$1.2\times10^{9}$ $\jpsi$ decays, in which the decays with measured
branching fractions are generated by {\sc EvtGen}~\cite{Ping2008} and
the remaining $\jpsi$ decays are generated according to the {\sc
  lundcharm}~\cite{Lundcharm} model. Simulated events are subject to the
same selection procedure applied to data. No significant peaking
background sources have been identified in the invariant mass spectrum
of $\ks\ks\eta$. Dominant backgrounds stem from
$\jpsi\ar\g\ks\ks\piz$ and $\jpsi\ar\g\ks\ks\piz\piz$. These
non-$\eta$ backgrounds are considered in the partial wave analysis
(PWA) by selecting events from data in the $\eta$ sideband regions
defined as $0.45<M_{\g\g}<0.48$~\GeV~and $0.60<M_{\g\g}<0.63$~\GeV,
and they account for about 2.5\% of the total number of events in the
$\eta$ signal region.

A PWA of events satisfying $M_{\ks\ks\eta}<2.8$~\GeV~and
$M_{\ks\ks}<1.1$~\GeV~is performed to determine the parameters of the
structure around 1.85 \GeV. These restrictions reduce complexities
due to additional intermediate processes. The signal amplitudes are
parameterized as sequential two-body decays, according to the isobar
model: $\jpsi\ar\g X,~X \ar Y\eta$ or $Z\ks$, where $Y$ and $Z$
represent the $\ks\ks$ and $\ks\eta$ isobars, respectively. Parity
conservation in the $\jpsi\ar\g\ks\ks\eta$ decay restricts the
possible $J^{PC}$ of the $\ks\ks\eta$ ($X$) system to be $0^{-+}$,
$1^{++}$, $2^{++}$,$2^{-+}$, $3^{++}$, etc. In this Letter, only spins
$J<3$ and possible $S$-wave or $P$-wave decays of the $X$ are
considered. The amplitudes are constructed using the covariant tensor
formalism described in Ref.~\cite{zoubs}. The relative magnitudes and
phases of the partial wave amplitudes are determined by an unbinned
maximum likelihood fit to data. The contribution of non-$\eta$
background events is accounted for in the fit by subtracting the
negative log-likelihood (NLL) value obtained for events in the $\eta$
sideband region from the NLL value obtained for events in the $\eta$
signal region. The statistical significance of a contribution is
estimated by the difference in NLL with and without the particular
contribution, taking the change in degrees of freedom into account.

\begin{figure*}[htpb]
 \centering
   \includegraphics[width=5.8cm,height=5.8cm]{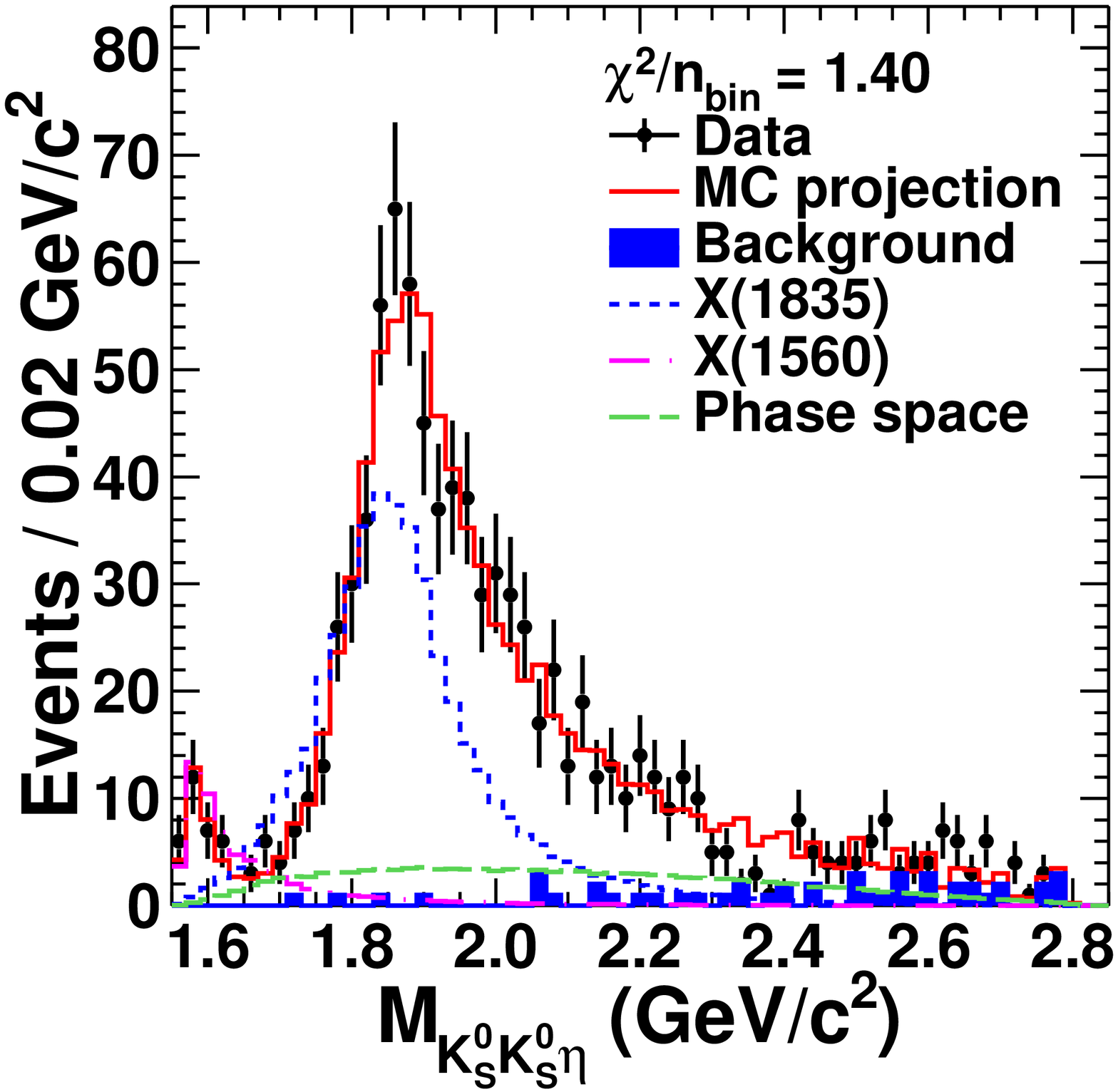}
   \put(-132,152){(a)}
   \includegraphics[width=5.8cm,height=5.8cm]{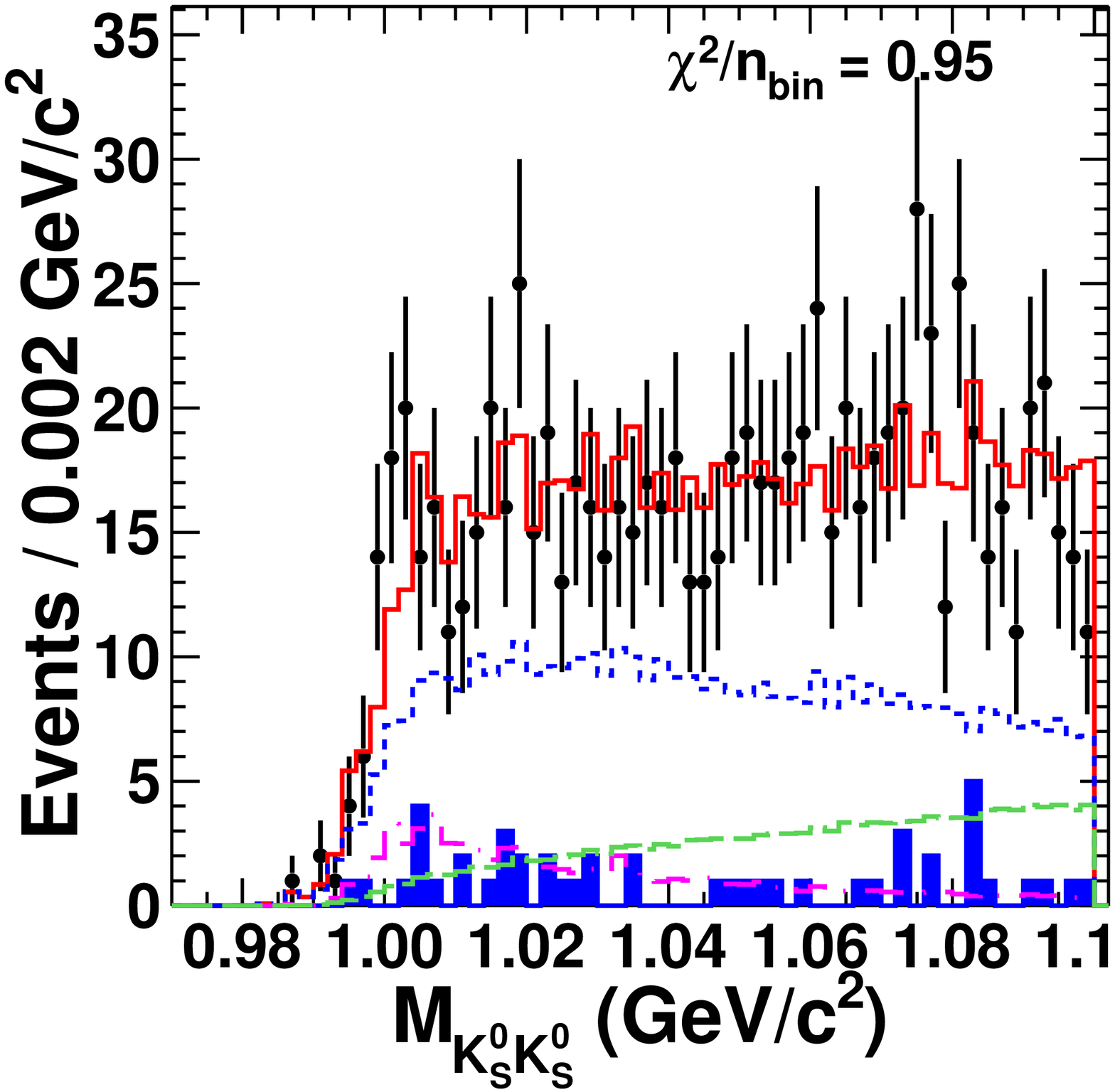}
   \put(-132,152){(b)}
   \includegraphics[width=5.8cm,height=5.8cm]{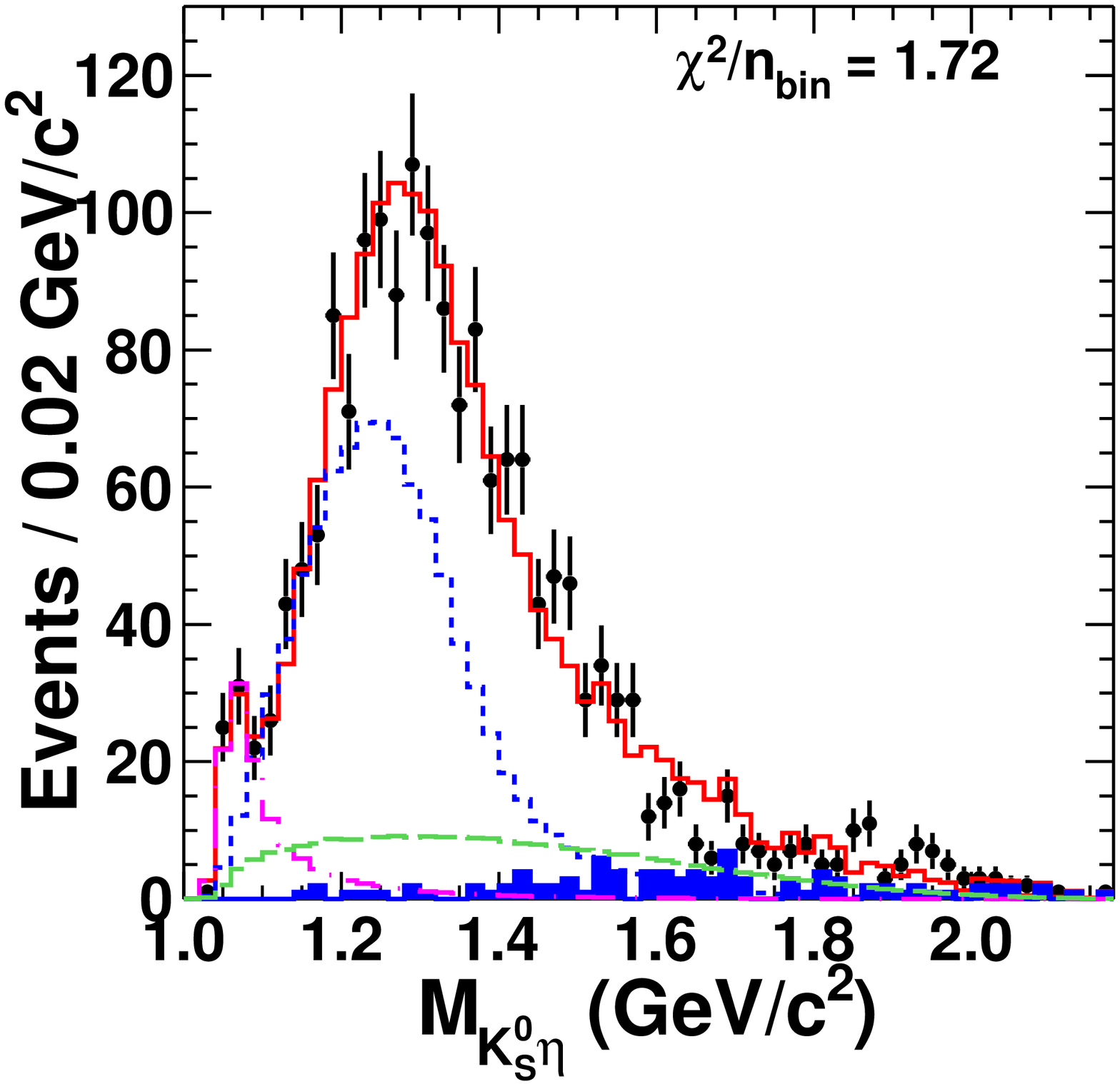}
   \put(-132,152){(c)}\\
   \includegraphics[width=5.8cm,height=5.8cm]{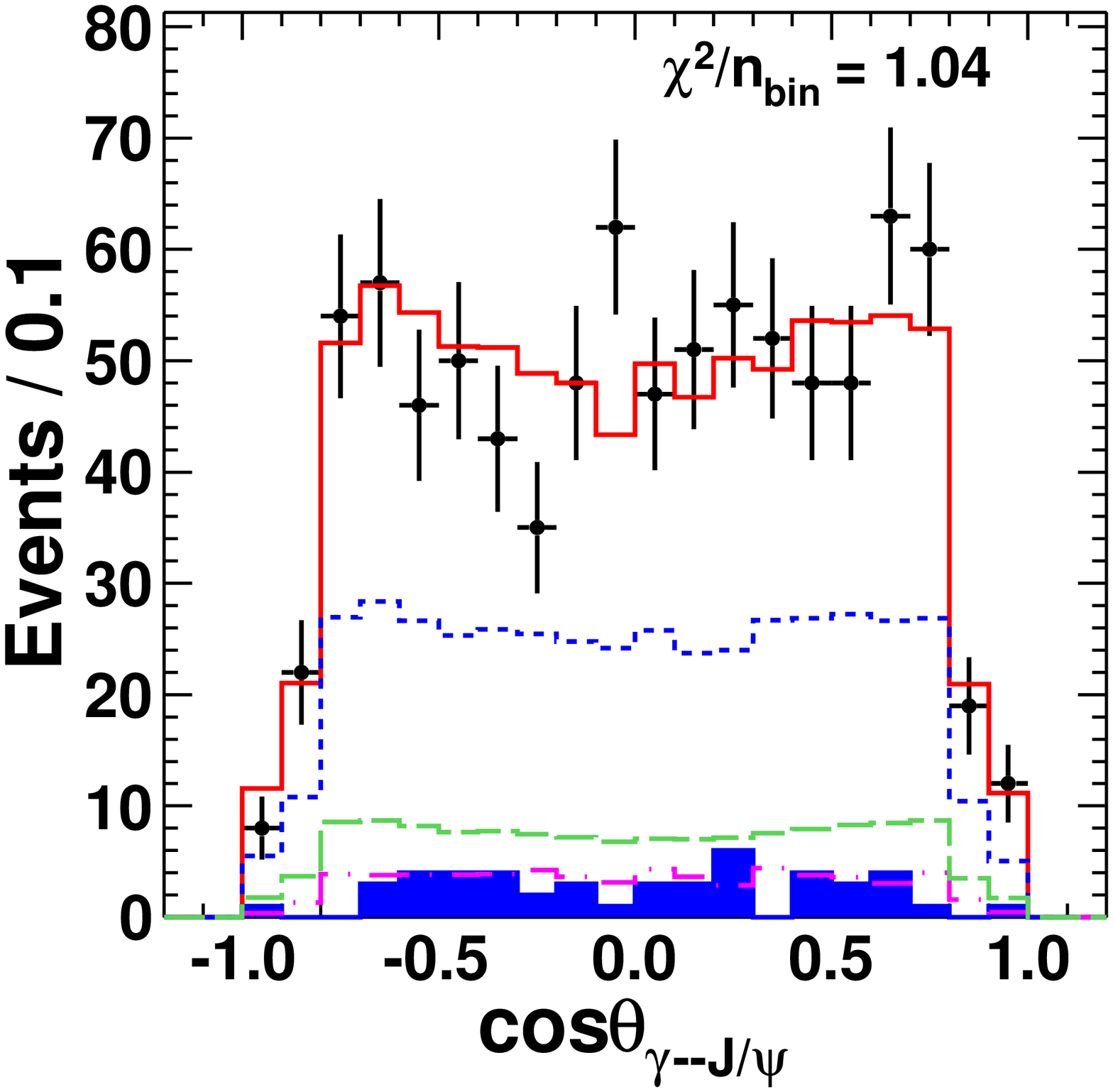}
   \put(-132,152){(d)}
   \includegraphics[width=5.8cm,height=5.8cm]{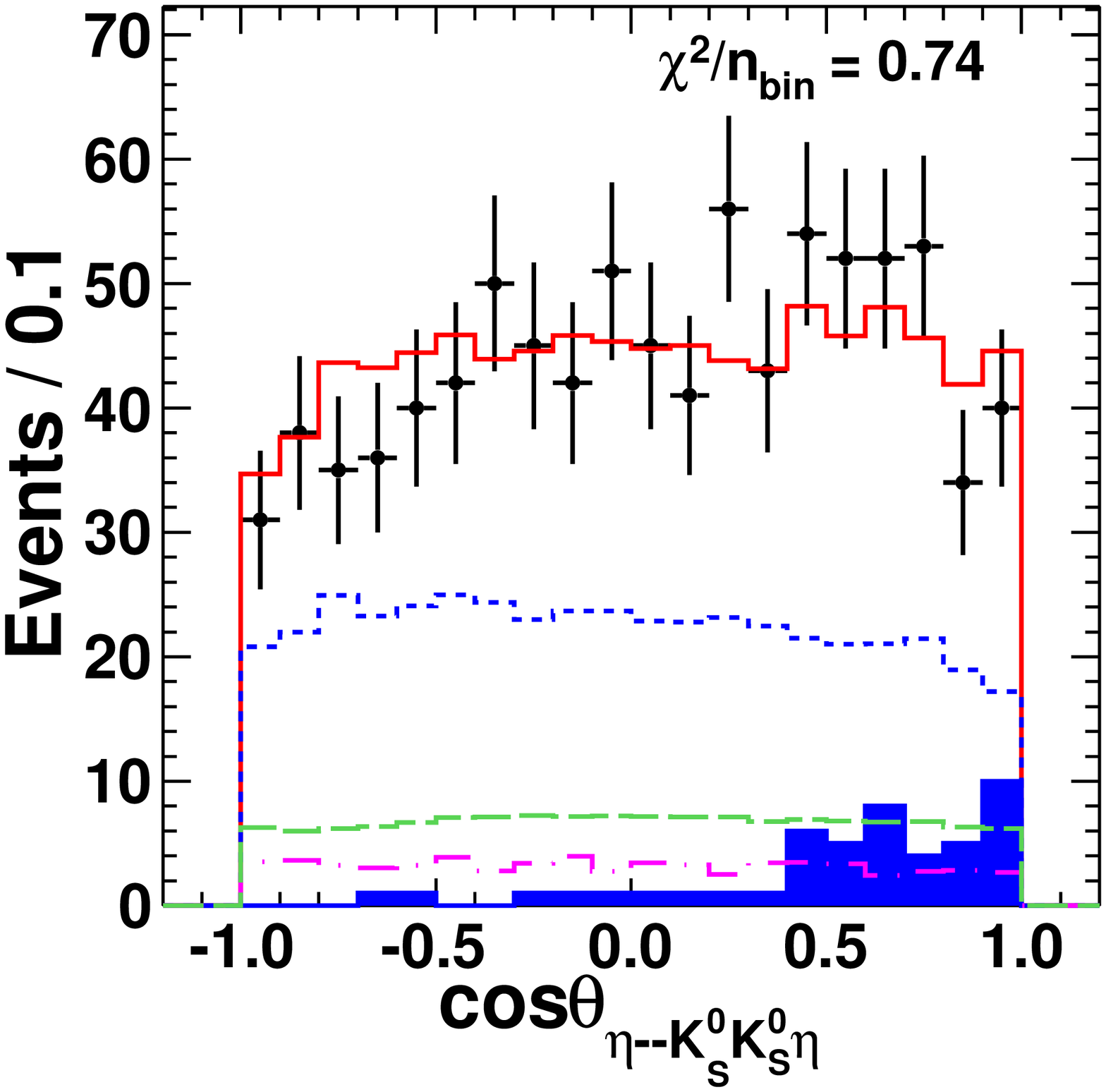}
   \put(-132,152){(e)}
   \includegraphics[width=5.8cm,height=5.8cm]{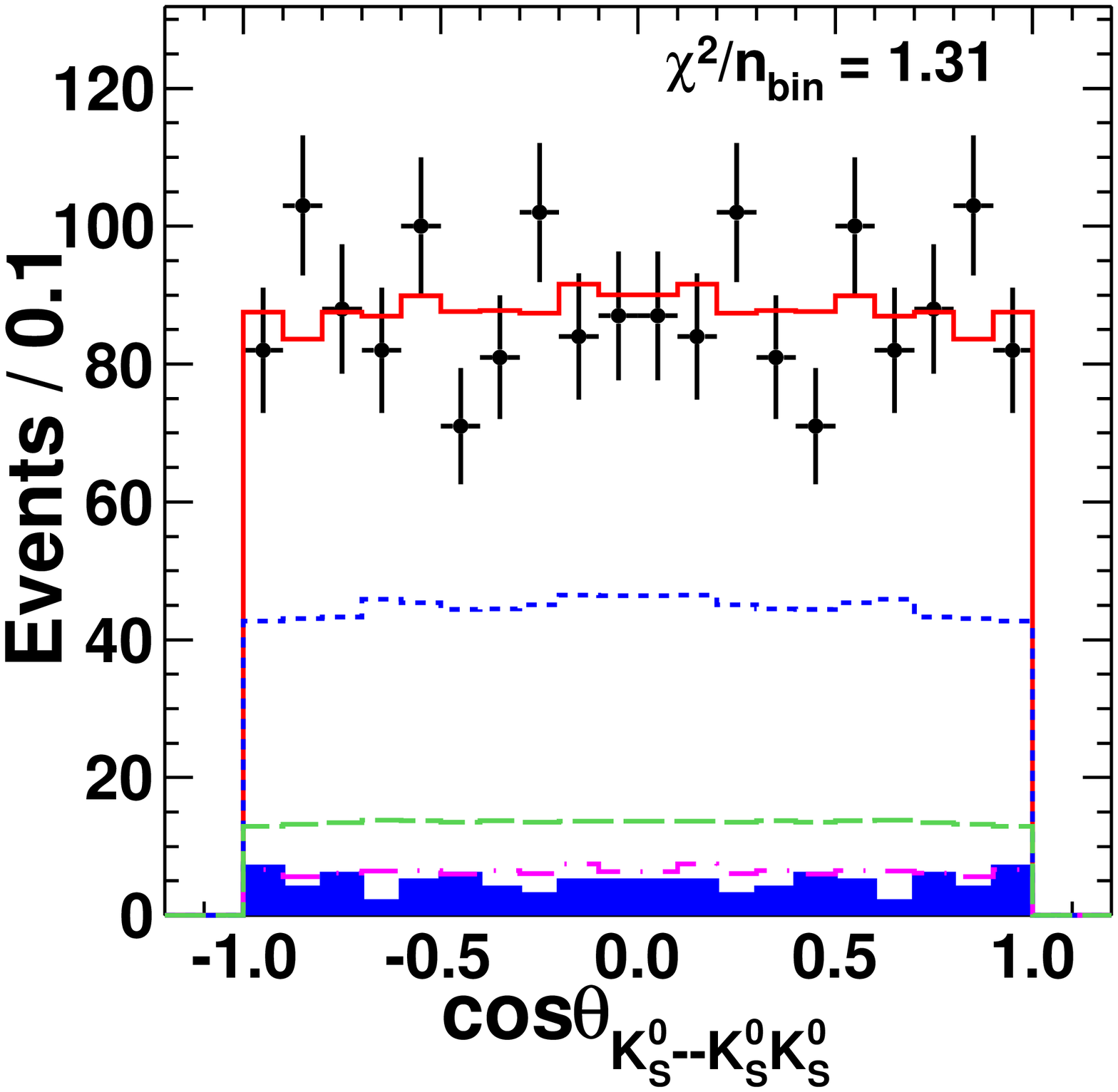}
   \put(-132,152){(f)}
 \caption{\label{proj_pwa_optsol}
   Comparisons between data and PWA fit projections. (a),
   (b), and (c) are the invariant mass distributions of $\ks\ks\eta$, $\ks\ks$, and
   $\ks\eta$ (two entries/event), respectively. (d)-(f) are the angular distributions of
   $\cos\theta$, where $\theta$ is the polar angle of (d) $\g$ in the $\jpsi$ rest system;
   (e) $\eta$ in the $\ks\ks\eta$ rest system; and (f) $\ks$ in the $\ks\ks$ rest system
   (two entries/event). The dots with error bars are data, the solid histograms are the
   PWA total projections, the shaded histograms are the non-$\eta$ backgrounds estimated
   by the $\eta$ sideband, and the short-dashed, dash-dotted, and long-dashed histograms
   show the contributions of $X(1835)$, $X(1560)$, and the nonresonant component,
   respectively.}
\end{figure*}

Our initial PWA fits include an $X(1835)$ resonance in the
$f_{0}(980)\eta$ channel and a nonresonant component in one of the
possible decay channels $f_{0}(980)\eta$, $f_{0}(1500)\eta$ or
$f_{2}(1525)\eta$. All possible $J^{PC}$ combinations of the $X(1835)$
and the nonresonant component are tried. We then extend the fits by
including an additional resonance at lower $\ks\ks\eta$ mass. This
additional component, denoted here as the $X(1560)$, improves the fit
quality when it is allowed to interfere with the $X(1835)$. Our final
fits show that the data can be best described with three components:
$X(1835)\ar\fz\eta$, $X(1560)\ar\fz\eta$, and a nonresonant
$f_{0}(1500)\eta$ component. The $J^{PC}$ of the $X(1835)$, the
$X(1560)$, and the nonresonant component are all found to be $0^{-+}$.
The $X(1835)$, $X(1560)$, and $f_{0}(1500)$ are described by
nonrelativistic Breit-Wigner functions, where the intrinsic widths
are not energy dependent. The masses and widths of the $X(1835)$ and
$X(1560)$ are derived by scanning each over a certain range. The
$f_{0}(1500)$ mass and width are fixed to the values reported in
Ref.~\cite{pdg2014}. The $\fz$ is parameterized by the Flatt\'e
formula~\cite{flatte}, with the parameters fixed to the
values reported by BESII~\cite{f0980param_bes2} in the channels
$\jpsi\ar\phi\pip\pim$ and $\jpsi\ar\phi\kap\kam$. The scan returns a mass
and width of the $X(1835)$ of $1844\pm9$\,\MeV~and
$192^{+20}_{-17}$\,MeV, respectively. The mass and width of the
$X(1560)$ are determined to be $1565\pm8$\,\MeV~and
$45^{+14}_{-13}$\,MeV, respectively. Using a detection efficiency of
5.5\%, obtained by a MC sample weighted by partial wave amplitudes,
the product branching fraction of $\jpsi\ar\g X(1835)$ and
$X(1835)\ar\ks\ks\eta$ ($\mathcal{B}_{X(1835)}$) is calculated to be
$(3.31^{+0.33}_{-0.30})\times 10^{-5}$, where the decay
$X(1835)\ar\ks\ks\eta$ is dominated by $f_{0}(980)$ production. The
$\ks\ks\eta$, $\ks\ks$, $\ks\eta$ mass spectra and the distributions
of the $\jpsi$, $\ks\ks\eta$ and $\ks\ks$ decay angles are shown in
Fig.~\ref{proj_pwa_optsol}. Overlaid on the data are the PWA fit
projections, as well as the individual contributions from each
component. The $\chi^2/n_\text{bin}$ value is displayed on each figure to
demonstrate the goodness of fit. We evaluate the significance by applying
the likelihood ratio test, performing a separate fit for every systematic
variation detailed below.
The most conservative statistical significances of
the $X(1835)$ and $X(1560)$ are $12.9\sigma$ and $8.9\sigma$,
respectively.

Various fits are performed by changing the $J^{PC}$ and decay mode of
the nonresonant component compared to the nominal solution described
above. The NLL value of a fit with a $1^{++}$ nonresonant
$f_{0}(1500)\eta$ component is only worse by 0.8 compared to the
nominal solution, which indicates that we cannot distinguish between
the two spin assignments of the nonresonant component with our
present statistics. This ambiguity introduces large systematic
uncertainties in the $\mathcal{B}_{X(1835)}$, since the interference
between the $X(1835)$, $X(1560)$, and the nonresonant component
depends on the spin assignment of the latter. To establish the
$J^{PC}$ of the $X(1835)$, we perform a series of PWA fits assuming
alternative $J^{PC}$ hypotheses for both the $X(1835)$ and the
nonresonant contribution. For the nonresonant contribution, we also
test several possible decay channels [$f_{0}(980)\eta$,
$f_{0}(1500)\eta$, and $f_{2}(1525)\eta$] in turn. For each
nonresonant component assumption, the $X(1835)$ $0^{-+}$ hypothesis
is significantly better than the $1^{++}$ or $2^{-+}$ hypotheses, with
the NLL value improving by at least 41.6 units. Analogously, we
perform the same series of PWA fits for the $X(1560)$. Again the
$0^{-+}$ hypothesis for the $X(1560)$ always yields a significantly
better fit result than other $J^{PC}$ assignments, with the NLL value
improving by at least 12.8 units.

We evaluate the contributions from additional well-known resonances by
adding them individually to the fit. We consider all possible
combinations for $X$ and its subsequent decay products $Y$ and $Z$ as
given in Ref.~\cite{pdg2014}: for $X$, this includes $\eta(1760)$,
$\eta(2225)$, $f_1(1510)$, $\eta_2(1870)$, $f_2(1810)$, $f_2(1910)$,
$f_2(1950)$, $f_2(2010)$, $f_2(2150)$, $f_2(2300)$, $f_2(2340)$,
$f_J(2220)$; for $Y$, $f_{0}(980)$, $f_{0}(1500)$, $f_{0}(1710)$,
$f_{2}(1270)$ and $f_{2}(1525)$; for $Z$, $K^{*}(1410)$,
$K^{*}(1680)$, $K^{*}_{0}(1430)$, $K^{*}_{0}(1950)$, $K^{*}_{2}(1430)$,
and $K^{*}_{2}(1980)$. Additional nonresonant contributions with
various $J^{PC}$ and decay modes are studied as well. The statistical
significances of the additional contributions are smaller than
5$\sigma$. In order to check the possible contribution from a
nonresonant $\ks\ks$ process, we add a $X(1835)\ar(\ks\ks)_{S}\eta$
process into the nominal solution, where $(\ks\ks)_{S}$ refers to a
nonresonant $\ks\ks$ contribution in a relative $S$ wave. We find the
resulting significance of the $X(1835)\ar f_{0}(980)\eta$ process and
the $X(1835)\ar(\ks\ks)_{S}\eta$ process to be 6.8$\sigma$ and
1.6$\sigma$, respectively, so we do not include the latter process in
the nominal solution. We also test a fit by changing the decay mode
of the $X(1560)$ in the nominal solution from $f_{0}(980)\eta$ to
$(\ks\ks)_{S}\eta$; the fit with $X(1560)\ar(\ks\ks)_{S}\eta$ has
almost the same fit quality as the nominal solution. Therefore, with
the present statistics, we cannot draw a conclusion about the
$X(1560)$ decay mode. The largest differences in masses and widths of
the $X(1835)$ and $X(1560)$ and the product branching fraction
$\mathcal{B}_{X(1835)}$ between all above alternative fits and the
nominal solution are taken as systematic uncertainties
from the components in the nominal solution.

For the measurements of the masses and widths of the $X(1835)$ and
$X(1560)$ and the product branching fraction $\mathcal{B}_{X(1835)}$,
we include the following sources of systematic uncertainties in
addition to the sources discussed above: we change the $\ks\ks$ mass
range to $M_{\ks\ks}<$ 1.05, 1.15 and 1.20~\GeV; we change the $\fz$
mass and coupling constants in the Flatt\'e formula to other
experimental
measurements~\cite{f0980param_SND,f0980param_KLOE,f0980param_BNL_zoubs};
we change the $f_{0}(1500)$ mass and width by one standard
deviation~\cite{pdg2014}; we increase and decrease the non-$\eta$
background level by one standard deviation; we change the
parameterization of the $X(1835)$ and $X(1560)$ line shape to a
Breit-Wigner function whose intrinsic width is
energy-dependent~\cite{BW_s}; and we replace the $X(1560)$ by
$\eta(1405)$ or $\eta(1475)$. For the systematic errors of the
product branching fraction $\mathcal{B}_{X(1835)}$, we also consider
the following additional uncertainties. The $\ks$ reconstruction
efficiency is studied using two control samples of
$\jpsi\ar K^{*\pm}K^{\mp}$ and $\jpsi\ar\phi\ks K^{\pm}\pi^{\mp}$,
while the photon detection efficiency is investigated based on a clean
sample of $\jpsi\ar\rho\pi$. The differences between data and MC
simulation are 1.0\% for each $\ks$ and 1.0\% for each
photon~\cite{sys_photon}. A control sample of $\jpsi\ar\g\ks\ks\piz$
is selected to estimate the uncertainty associated with the $4C$
kinematic fit. The efficiency is the ratio of the signal yields
with and without the kinematic fit requirement $\chi^{2}_{4C}<40$. The
difference between data and MC simulation, 1.5\%, is assigned as the systematic
uncertainty. We also consider the uncertainties from the number of
$\jpsi$ events~\cite{njpsi_09,njpsi_all} and the branching fractions
of $\ks\ar\pip\pim$ and $\eta\ar\g\g$~\cite{pdg2014}. We change the
mass and width of $X(1835)$ or $X(1560)$ by 1 standard deviation of
the statistical uncertainty. The individual
uncertainties are assumed to be independent and are added in
quadrature to obtain the total systematic uncertainties as presented
in the Supplemental Material~\cite{supplemental_material}.

In summary, a PWA of $\jpsi\ar\g\ks\ks\eta$ has been performed in the
mass range $M_{\ks\ks\eta}<2.8$~\GeV~after requiring
$M_{\ks\ks}<1.1$~\GeV. The PWA fit requires a contribution from
$X(1835)\ar\ks\ks\eta$ with a statistical significance greater than
12.9$\sigma$, where the $X(1835)\ar\ks\ks\eta$ is dominated by
$f_{0}(980)$ production. The spin parity of the $X(1835)$ is
determined to be $0^{-+}$. The mass and width of the $X(1835)$ are
measured to be $1844\pm9(\text{stat})^{+16}_{-25}(\text{syst})$\,\MeV~and
$192^{+20}_{-17}(\text{stat})\phantom{}^{+62}_{-43}(\text{syst})$\,MeV,
respectively. The corresponding product branching fraction
$\mathcal{B}_{X(1835)}$ is measured to be
$(3.31^{+0.33}_{-0.30}(\text{stat})\phantom{}^{+1.96}_{-1.29}(\text{syst})) \times
10^{-5}$.
The mass and width of the $X(1835)$ are consistent with the values
obtained from the decay $\jpsi\ar\g\pip\pim\eta^{\prime}$ by
BESIII~\cite{x1835_bes3}. These results are all first-time
measurements and provide important information to further
understand the nature of the $X(1835)$.

Another $0^{-+}$ state, the $X(1560)$, also is observed in data with a
statistical significance larger than 8.9$\sigma$ and is seen to interfere with
the $X(1835)$. The mass and width of the $X(1560)$ are determined to be
$1565\pm8(\text{stat})^{+0}_{-63}(\text{syst})$\,\MeV~and
$45^{+14}_{-13}(\text{stat})\phantom{}^{+21}_{-28}(\text{syst})$\,MeV, respectively.
The mass and width of the $X(1560)$ are consistent with those of the
$\eta(1405)$ and $\eta(1475)$ as given in Ref.~\cite{pdg2014} within
$2.0\sigma$ and $1.4\sigma$, respectively.
Present statistics do not allow us to conclusively determine if the
$X(1560)$ is the same state as the $\eta(1405)$/$\eta(1475)$ or a new meson.
More statistics in this analysis and an amplitude analysis of $\jpsi\ar\g\eta\pi^{0}\pi^{0}$ and
$\jpsi\ar\g\ks\ks\pi^{0}$ processes may help to understand the nature of the $X(1560)$.

The BESIII collaboration thanks the staff of BEPCII and the IHEP
computing center for their strong support. This work is supported in
part by National Key Basic Research Program of China under Contract
No.~2015CB856700; National Natural Science Foundation of China (NSFC)
under Contracts No.~11125525, No.~11235011, No.~11322544, No.~11335008, No.~11425524;
the Chinese Academy of Sciences (CAS) Large-Scale Scientific Facility
Program; the CAS Center for Excellence in Particle Physics (CCEPP);
the Collaborative Innovation Center for Particles and Interactions
(CICPI); Joint Large-Scale Scientific Facility Funds of the NSFC and
CAS under Contracts No.~11179007, No.~U1232201, No.~U1332201; CAS under
Contracts No.~KJCX2-YW-N29, No.~KJCX2-YW-N45; 100 Talents Program of CAS;
INPAC and Shanghai Key Laboratory for Particle Physics and Cosmology;
German Research Foundation DFG under Contract No.~Collaborative
Research Center CRC-1044; Istituto Nazionale di Fisica Nucleare,
Italy; Ministry of Development of Turkey under Contract
No.~DPT2006K-120470; Russian Foundation for Basic Research under
Contract No.~14-07-91152; U.S.~Department of Energy under Contracts
No.~DE-FG02-04ER41291, No.~DE-FG02-05ER41374, No.~DE-FG02-94ER40823,
No.~DESC0010118; U.S.~National Science Foundation; University of Groningen
(RuG) and the Helmholtzzentrum fuer Schwerionenforschung GmbH (GSI),
Darmstadt; WCU Program of National Research Foundation of Korea under
Contract No.~R32-2008-000-10155-0.


\end{document}